\pdfoutput=1

\documentclass[11pt]{article}

\usepackage[]{EMNLP2022}

\usepackage{times}
\usepackage{latexsym}

\usepackage[T1]{fontenc}

\usepackage[utf8]{inputenc}

\usepackage{microtype}

\usepackage{inconsolata}

\usepackage{threeparttable}
\usepackage{enumitem}
\usepackage{hyperref}
\usepackage{caption}
\usepackage{graphicx}
\usepackage{amsmath}
\usepackage{booktabs}
\usepackage{subcaption}

\title{TeleMelody: Lyric-to-Melody Generation with a Template-Based Two-Stage Method}

\author{Zeqian Ju\textsuperscript{\rm 1}, 
Peiling Lu\textsuperscript{\rm 2}, 
Xu Tan\Thanks{ This work was conducted at Microsoft. Corresponding author: Xu Tan, xuta@microsoft.com}\ \ \textsuperscript{\rm 2}, 
Rui Wang\textsuperscript{\rm 2}, 
Chen Zhang\textsuperscript{\rm 3}, \\ 
\textbf{Songruoyao Wu\textsuperscript{\rm 3},} 
\textbf{Kejun Zhang\textsuperscript{\rm 3},}
\textbf{Xiangyang Li\textsuperscript{\rm 1}, }
\textbf{Tao Qin\textsuperscript{\rm 2},  }
\textbf{Tie-Yan Liu\textsuperscript{\rm 2}}  \\
 \textsuperscript{\rm 1} University of Science and Technology of China\\
 \textsuperscript{\rm 2} Microsoft Research Asia, 
 \textsuperscript{\rm 3} Zhejiang University, China\\
\url{https://github.com/microsoft/muzic} 
}

\begin{document}
\maketitle
\begin{abstract}
 Lyric-to-melody generation is an important task in automatic songwriting. Previous lyric-to-melody generation systems usually adopt end-to-end models that directly generate melodies from lyrics, which suffer from several issues: 1) lack of paired lyric-melody training data; 2) lack of control on generated melodies. In this paper, we develop TeleMelody, a two-stage lyric-to-melody generation system with music template (e.g., tonality, chord progression, rhythm pattern, and cadence) to bridge the gap between lyrics and melodies (i.e., the system consists of a lyric-to-template module and a template-to-melody module). TeleMelody has two advantages. First, it is data efficient. The template-to-melody module is trained in a self-supervised way (i.e., the source template is extracted from the target melody) that does not need any lyric-melody paired data. The lyric-to-template module is made up of some rules and a lyric-to-rhythm model, which is trained with paired lyric-rhythm data that is easier to obtain than paired lyric-melody data. Second, it is controllable. The design of the template ensures that the generated melodies can be controlled by adjusting the musical elements in the template. Both subjective and objective experimental evaluations demonstrate that TeleMelody generates melodies with higher quality, better controllability, and less requirement on paired lyric-melody data than previous generation systems. 
\end{abstract}

\section{Introduction}
Music is a universal natural language for communication. With the rapid development of artificial intelligence, automatic songwriting has drawn much attention from both academia and industry. Automatic songwriting covers many tasks, such as lyric generation \cite{malmi2016dopelearning,xue2021deeprapper}, melody generation \cite{wu2020popmnet,choi2016text,zhu2018xiaoice}, lyric-to-melody generation \cite{yu2021conditional,sheng2020songmass,bao2019neural, lee-etal-2019-icomposer}, and melody-to-lyric generation \cite{sheng2020songmass,watanabe-etal-2018-melody,li-etal-2020-rigid}. In this paper, we focus on lyric-to-melody generation, since it is one of the most important and common tasks in songwriting and is still under-explored.
Recent years, deep learning techniques have been widely used to develop end-to-end lyric-to-melody systems \cite{yu2021conditional,sheng2020songmass,bao2019neural, lee-etal-2019-icomposer}. However, these systems suffer from the following issues: 1) They require large amount of paired lyric-melody data to learn the correlation between syllables in lyrics and notes in melodies \cite{sheng2020songmass}. However, collecting lots of paired data is quite difficult and costy. \citeauthor{sheng2020songmass} have attempted to alleviate the low-resource challenge by unsupervised pre-training on lyric-to-lyric and melody-to-melody models. However, the utilization of unpaired data helps on the understanding and generation of lyrics and melodies while has little effect on the correlation learning between lyrics and melodies. 2) They generate melodies directly from lyrics, which hinders end users to control musical elements (e.g. tonality and chord progression) over the generation. Without controllability, requirements from users may be ignored and the application scenarios are limited.

In this paper, we propose TeleMelody$\footnote{Tele is from TEmpLatE.}$, a two-stage lyric-to-melody generation system with a carefully designed template as a bridge to connect lyrics and melodies. The template contains tonality, chord progression, rhythm pattern, and cadence. This designed template is effective because: 1) It is convenient to be extracted from melodies and predicted from lyrics and can successfully catch their characteristics; 2) It is easy to be manipulated by users on demand. Accordingly, we break down the lyric-to-melody task into a lyric-to-template module and a template-to-melody module. This can reduce the task difficulty, improve data efficiency and achieve better controllability. The details are described as follows:

This two-stage framework can help reduce the difficulty of learning the correlation between lyrics and melodies. In the template-to-melody module, we train a template-to-melody model with templates extracted from melodies by rules. Generating melodies from templates is much easier than from lyrics, since the correlation between templates and melodies is much stronger than that between lyrics and melodies, and the paired template-melody data can be easily got from a self-supervised way. In the lyric-to-template module, rhythm pattern in template is obtained by a lyric-to-rhythm model, which is trained with paired lyric-rhythm data. This paired data is obtained by extracting the rhythm pattern from crawled lyric-audio data through audio processing tools, which is much easier to get than paired lyric-melody data. Cadence is inferred based on punctuation mappings. Chord progression and tonality in the template can be acquired with predefined musical knowledge. In this way, the two modules can rely on self-supervised learning or data mining on external lyric-audio data, which do not require any paired lyric-melody data and are more data-efficient than end-to-end models.

Moreover, benefiting from the template based framework, end users can control the generated melodies by changing the musical elements in templates. Besides, in the sequence-to-sequence based template-to-melody model, we use musical knowledge to guide the learning of attention alignments between the template tokens and the corresponding melody tokens, which can lead to better controllability.

The main contributions of this work are as follows:
\begin{itemize}
\item We propose TeleMelody, a two-stage lyric-to-melody system with a carefully designed template as the bridge. It decomposes lyric-to-melody generation into a lyric-to-template module and a template-to-melody module. This framework can help reduce the task difficulty and improve data efficiency.

\item Chord progression, tonality, rhythm pattern and cadence designed in templates can help control basic musical elements and high-level music structures. We introduce alignment regularization based on musical knowledge to ensure better controllability for generated melodies.

\item Experimental results demonstrate that TeleMelody significantly outperforms previous end-to-end lyric-to-melody generation models in terms of both objective and subjective evaluations on generation quality, and is capable of better controlling the generated melodies.
\end{itemize}

\section{Background}

\subsection{Lyric-to-Melody Generation}
Considerable development in lyric-to-melody generation has been seen in recent years, from rule-based or statistical methods to deep learning methods. Rule-based or statistical methods usually need lots of manual designs based on domain knowledge in music, and hinder end users to control musical elements. \citeauthor{monteith2012automatic} generate rhythm pattern based on rules, and construct an n-gram model to predict note pitch$\footnote{\url{https://en.wikipedia.org/wiki/Pitch\_(music)}}$. \citeauthor{long2013t} and \citeauthor{rabin1963probabilistic} learn the lyric-note correlation by performing statistical method with limited paired data. In these works, the melody is generated with the control of zero or just one specific musical element. \citeauthor{fukayama2010automatic} obtain the optimal pitch sequence by maximizing the conditional probability given chords, tonality, accompaniment bass, rhythm and pitch accent information, but the generated melodies may suffer from bad musical structure without repetition patterns. Meanwhile the algorithm cannot be directly applied to lyrics written in ``stress accent'' languages like English.

Recently, developing lyric-to-melody systems based on machine learning methods attracts lots of attentions.  \citeauthor{scirea2015smug} allocate the same number of notes as the syllables in lyrics using Markov chain. \citeauthor{ackerman2017algorithmic} leverage random forest model to construct a rhythm model and a melody model separately on paired lyric-audio data. \citeauthor{bao2019neural} and \citeauthor{yu2021conditional} use sequence-to-sequence models to generate melody from lyrics. However, deep learning methods usually require large amount of paired lyric-melody data for learning the correlation between lyrics and melodies. \citeauthor{sheng2020songmass} attempt to address low-resource challenge by performing pre-training for lyric-to-lyric and melody-to-melody modules, and incorporating supervised learning into the pre-training to learn a shared latent space between lyrics and melodies. But the challenge is not well addressed since the unpaired data has not been sufficiently utilized on correlation learning between lyrics and melodies. Moreover, these works do not consider controlling specific musical elements of generated melodies. 
In this paper, we propose TeleMelody, a two-stage template-based system, which consists of a lyric-to-template module and a template-to-melody module. The two-stage framework can help address the issues of limited paired data, and the designed template together with gularization in this framework is able to ensure better controllability over generated melodies.

\begin{figure}[t]
\centering
\subcaptionbox{The melody, lyric, and chord progression.\label{fig:melody}}{
\includegraphics[width=1.0\columnwidth]{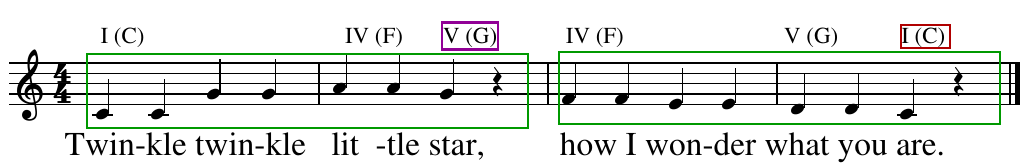}
}

\subcaptionbox{The corresponding template.\label{fig:template}}{
\includegraphics[width=1.0\columnwidth]{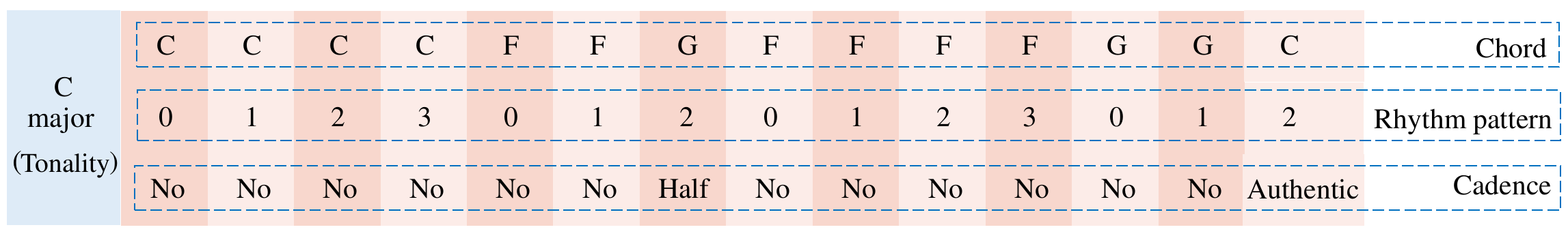}
}
\vspace{-3mm}
\caption{The song ``Twinkle Twinkle Little Star'' in ``C major'' tonality.}

\end{figure}

\subsection{Music Background Knowledge}
In this subsection, we use the song ``Twinkle Twinkle Little Star'' in Figure \ref{fig:melody} as an example to introduce musical elements in the template.

\begin{itemize}
  \item Tonality$\footnote{\url{https://en.wikipedia.org/wiki/Tonality}}$ is composed of a scale$\footnote{\url{https://en.wikipedia.org/wiki/Scale\_(music)}}$ and a root note. For example, the tonality of the melody in Figure \ref{fig:melody} is ``C major'', since notes are ordered within pitches in major scale with the root pitch ``C''.
  
  \item Chord progression$\footnote{\url{https://en.wikipedia.org/wiki/Chord\_progression}}$ is an ordered sequence of chords. As in Figure \ref{fig:melody}, the chord progression is ``I-IV-V-IV-V-I'' (``C-F-G-F-G-C'' in C major scale). Chord progression, interacting with melody, should create a sense of harmony when composing music.

  \item Rhythm$\footnote{\url{https://en.wikipedia.org/wiki/Rhythm\#Composite\_rhythm}}$ refers to the pattern of occurrence of notes and rests. In Figure \ref{fig:melody}, each note is aligned with a syllable and notes in green boxes are in the same rhythm patterns.

  \item Cadence$\footnote{\url{https://en.wikipedia.org/wiki/Cadence}\label{cadence}}$ occurs at the end of phrase and gives a sense of ending in melody, and is often aligned with punctuation marks in lyrics.
  In Figure \ref{fig:melody}, we assign the half cadence$\textsuperscript{\ref{cadence}}$ and the authentic cadence$\textsuperscript{\ref{cadence}}$ to the comma and the period respectively.
\end{itemize}

\section{Methodology}
Figure \ref{fig:overview} describes the lyric-to-melody generation system architecture connecting a lyric-to-template module and a template-to-melody module with a template. In this section, we introduce each component (i.e., template, lyric-to-template module, and template-to-melody module) in detail.

\subsection{Template}
\label{sec:template}
In this subsection, we introduce template in respects of definition, design principles, contained musical elements, and the connection with the generated melodies.

\begin{figure}[t!]
\centering
\subcaptionbox{Pipeline of TeleMelody.\label{fig:overview}}{

\includegraphics[width=0.8\columnwidth]{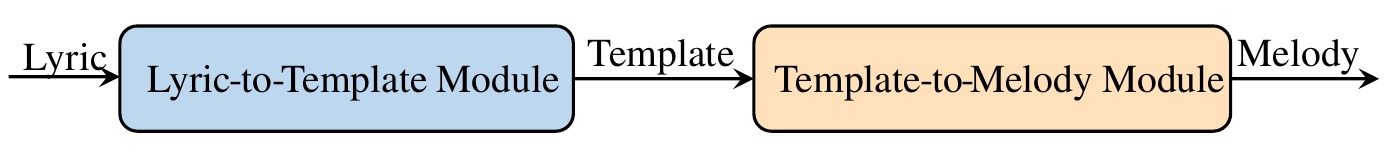}
}
\subcaptionbox{Architecture of lyric-to-template module.\label{fig:l2s}}{
\includegraphics[width=0.8\columnwidth]{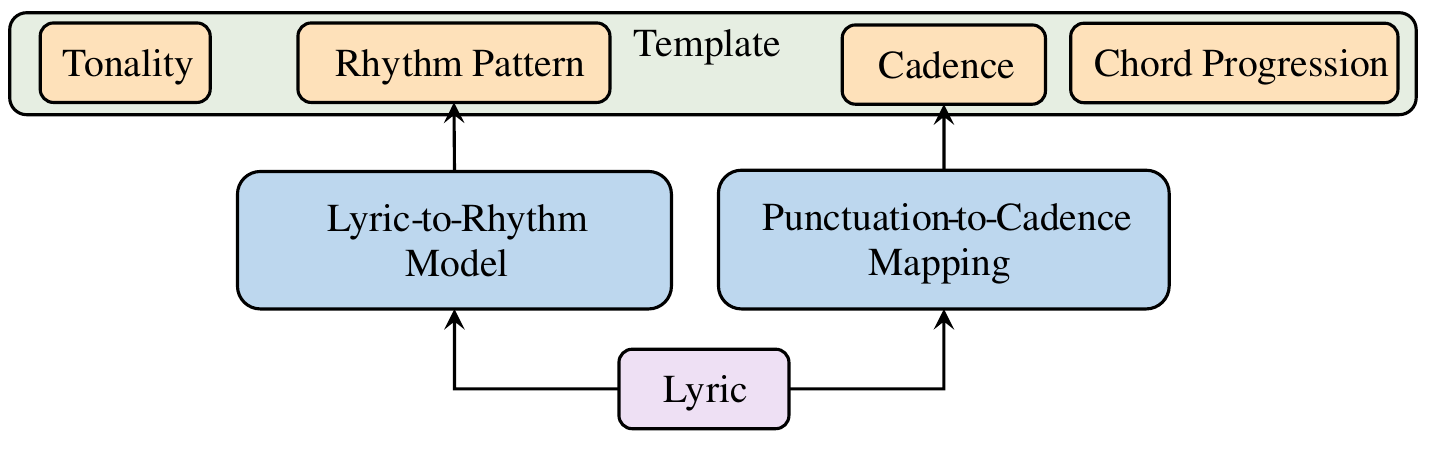}
}

\subcaptionbox{The template and its connection with melody.\label{fig:align}}{
\includegraphics[width=0.9\columnwidth]{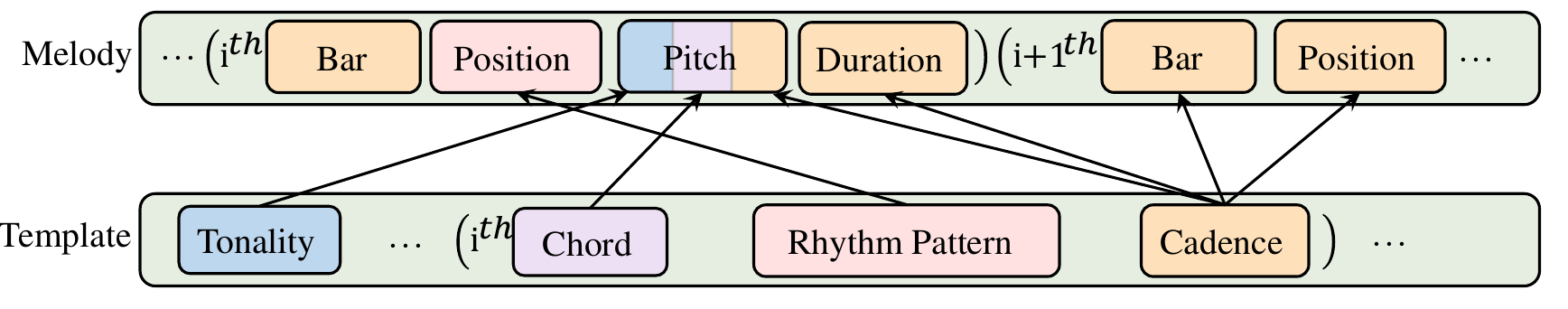}
}
\vspace{-3mm}
\caption{Architecture of TeleMelody.}
\end{figure}

The template is a well-designed sequence of musical elements that can capture the common attributes of lyrics and melodies. With the connection of this template, we decompose lyric-to-melody generation into a lyric-to-template module and a template-to-melody module as shown in Figure \ref{fig:overview}.

Concerning task characteristics, we propose following high-level principles for template design: 1) Templates can be extracted directly from melodies so that a template-to-melody model can be trained in a self-supervised way. 2) Templates obtained from lyrics should be in accordance with those extracted from melodies in order to bridge lyrics and melodies in inference. 3) Compared with hidden representations in end-to-end models, templates should be more easily manipulated on demand to achieve better controllability.

Based on above principles, the designed template consists of tonality, chord progression, rhythm pattern, and cadence. The template representation is shown in Figure \ref{fig:align}: we use one token at the start of the sequence to represent the tonality and three consecutive tokens (chord, rhythm pattern, cadence) to represent musical elements of each note. Figure \ref{fig:template} provides the example of template corresponds to the melody in Figure \ref{fig:melody}.

The influence of musical elements in templates on generated melodies are described as follows (shown in Figure \ref{fig:align}): 1) Tonality can control pitch distribution. 2) Chord can influence the harmony of generated melodies. 3) Rhythm pattern can constrain note position and control high-level musical structure with repetitive patterns. 4) Cadence can guarantee the accordance between punctuation in lyrics and note onset intervals in melodies.

\subsection{Template-to-Melody Module}
The template-to-melody module is to generate melodies from given templates with an encoder-attention-decoder Transformer model in a self-supervised way. In this subsection, we first introduce how to extract musical elements in templates from melodies for model training and then introduce alignment regularization, which can leverage musical knowledge to improve model controllability.

\paragraph{Template Extraction Method}
\label{sec:retrieval}
We introduce the method for extracting templates (i.e., tonality, chord progression, rhythm pattern, cadence) from melodies: 1) Tonality can be inferred according to the note pitch distribution of a whole melody, following \citeauthor{liang2020pirhdy}. 2) Chord progression can be inferred based on note pitch distribution through the Viterbi algorithm proposed by Magenta$\footnote{\url{https://github.com/magenta/note-seq/blob/master/note\_seq/chord\_inference.py}}$. 3) Rhythm pattern can be inferred based on position information of the note. 4) Cadence can be inferred by rules based on note pitch, onset interval and duration. We classify it in three classes: ``no cadence'', ``authentic cadence'', and ``half cadence''. Detailed rules are described in Appendix \ref{sec:cad}.


\paragraph{Alignment Regularization}
\label{sec:align}
With the introduction of the template, we can control melody generation by adjusting musical elements in the template. To further increase model controllability, we introduce musical knowledge to the template-to-melody model through well-designed alignment regularization \cite{garg2019jointly} during training. Each designed alignment is imposed on the encoder-decoder attention. This musical knowledge based design can provide the guidance for the model to learn the interpretable alignments between templates and melodies, which can help for better controllability. 

We denote $m_k$ as the $k^{th}$ note information token in melody sequence, and $t_j$ as the $j^{th}$ musical element token in template sequence. $\hat{w}$ denotes a $0$-$1$ matrix such that $\hat{w}_{k,j}=1$ if $m_k$ is aligned with $t_j$. We simply normalize the rows of matrix $\hat{w}$ to get a matrix $w$. We expect encoder-decoder attention weight $A_{k,j}$ between $m_k$ and $t_j$ to be closer to $w_{k,j}$ defined as:
\begin{equation}
\label{eq:w}
w_{k,j} =
\begin{cases}
\frac{1}{T}& \text{if $m_k$ is aligned to $t_j$,}\\
0& \text{otherwise,}
\end{cases} 
\end{equation}
\noindent where $T$ is the number of tokens in the template sequence that $m_k$ is aligned to.

We denote $J$ and $K$ as the number of
tokens in the source and target sentence respectively. The alignment regularization term is 
\begin{equation}
    L_{attn} = \frac{1}{K\times J}\sum_{k=1}^{K}\sum_{j=1}^{J} w_{k,j} \ \textit{log} A_{k,j}
\end{equation}
The overall loss is
\begin{equation}
    L = L_{nll} + \lambda_{attn}\ L_{attn}
\end{equation}  
where $L_{nll}$ is the negative log likelihood loss and $\lambda_{attn}$ is a hyperparameter. 

We give an example  in Figure \ref{fig:align} to illustrate the alignment between the consecutive $i^{th}$ and $i$+$1^{th}$ note in the melody, and the corresponding musical elements in the template. The note information is consisted of bar index, position in a bar, pitch and duration. The alignments mentioned in Equation (\ref{eq:w}) are designed as follows:
\begin{itemize}
  \item We add an alignment between tonality and every note pitch in the melody, since tonality controls the pitch distribution of the entire melody.
  
  \item We add an alignment between the chord of $i^{th}$ note in the template and pitch of $i^{th}$ note in the melody, since chord influences the note pitch.
  
  \item We add an alignment between the rhythm pattern of $i^{th}$ note in the template and the position of $i^{th}$ note in the melody, since rhythm pattern determines the note position.
  

    \item We add an alignment between the cadence of $i^{th}$ note in the template with the duration and pitch of $i^{th}$ note in the melody for their close relationship. Besides, we add alignments between cadence of $i^{th}$ note with the bar index and position of $i$+$1^{th}$ note, since onset intervals between adjacent notes can help distinguish ``no cadence'' from others.
\end{itemize}

\subsection{Lyric-to-Template Module}
In this subsection, we describe the lyric-to-template module, which generates musical elements (i.e., tonality, chord progression, rhythm pattern, and cadence) in templates from lyrics. Tonality and chord progression are weakly correlated with lyrics and can be manipulated on demand. Therefore, we focus on generating rhythm pattern and cadence in the following paragraphs.

\paragraph{Lyric-to-Rhythm Model}
We introduce a transformer based lyric-to-rhythm model to predict rhythm patterns in the template with given lyrics in an auto-regressive way. To collect adequate lyric-rhythm data for training, we extracted the paired data from crawled lyric-audio data with audio processing tools. The detailed pipeline is similar to \citeauthor{xue2021deeprapper} and is described in Appendix \ref{sec:l2r}. 	

\paragraph{Punctuation-Cadence Mapping}
Since punctuation marks in lyrics are closely related to cadences in melodies, we design musical knowledge based mapping as follows: 1) We align ``authentic cadence'' and ``half cadence” with period and comma in lyrics respectively for indicating the end of the sentence. 2) We align the ``no cadence” label with each syllable in lyrics since it is a voiced part. 

\section{Experimental Settings}
    
\subsection{Dataset}    
\label{sec:dataset}
We conduct experiments on both English (EN) and Chinese (ZH) lyric-to-melody generation tasks to evaluate our model. For lyric-to-template module, we collect paired lyric-rhythm data ($9,761$ samples in English and $74,328$ samples in Chinese) following the procedure in Appendix \ref{sec:l2r}. As for template-to-melody module, we obtain and process $45, 129$ midi data from a widely used dataset in music generation, LMD-matched MIDI dataset$\footnote{\url{https://colinraffel.com/projects/lmd}}$ \cite{raffel2016learning}, since the musical elements in designed template do not have much correlation to languages. The majority of the dataset is Pop/Rock, following with Electronic and Country \cite{ferraro2018large}. We provide detailed melody pre-processing in Appendix \ref{sec:process}. Statistics of lyric-template and template-melody dataset are shown respectively in Table \ref{tab:l2r} and Table \ref{tab:t2m} in Appendix \ref{sec:data_stat}. The code and models can be found at our project$\footnote{\url{https://github.com/microsoft/muzic}}$.


\subsection{System Configurations}
Both two Transformer models in lyric-to-template and template-to-melody module use the settings of $4$ encoder layers, $4$ decoder layers, $256$ hidden size and $4$ attention heads. The dropout is $0.2$ in lyric-to-template model and $0.0005$ in template-to-melody model. We use Adam optimizer with the learning rate of $0.0005$ in both models. The alignment regularization weight $\lambda_{attn}$ in template-to-melody model is set as $0.05$. To ensure the diversity of generated melodies, we use stochastic sampling inference following \citeauthor{huang2018music}. The temperature and top-$k$ parameters are set as $0.5$ and $2$ in lyric-to-rhythm generation, and as $0.5$ and as $10$ in template-to-melody generation.

\subsection{Evaluation Metrics} 

\paragraph{Objective Evaluation}
Following \citeauthor{sheng2020songmass}, we consider two metrics to measure the similarity of the generated and the ground-truth melodies: Similarity of pitch and duration distribution (PD and DD) and Melody distance (MD). Besides, we use accuracy of tonality, chord, rhythm pattern, and cadence (TA, CA, RA, AA) to measure the consistency between the generated melody and musical elements in the template. The more consistent the generated melody is with the template, the more controllable the model is. For ground-truth melodies, TA, RA, AA are $100\%$, while CA  is less than $100\%$ since introducing notes outside chord are encouraged to avoid monotony in songwriting. Therefore, CA scores is better if it is closer to that of the ground-truth melodies. Accordingly, we consider that the closer TA, CA, RA, and AA are to the ground-truth melodies, the more controllable the model is. The definitions of these metrics are described in Appendix \ref{sec:def}.

\paragraph{Subjective Evaluation} 
We invite $10$ participants (including 7 amateurs and 3 professionals) as human annotators to evaluate $10$ songs in each language. We require each participant to rate properties of the melodies in a five-point scale, from 1 (Poor) to 5 (Perfect). The whole evaluation is conducted in a blind-review mode. Referring to previous works \cite{sheng2020songmass,zhu2018xiaoice,watanabe-etal-2018-melody}, we use following subjective metrics to evaluate the generated melodies: 1) Harmony: Is the melody itself harmonious? 2) Rhythm : Is the rhythm  sounds natural and suitable for lyrics? 3) Structure: Does the melody consist of repetitive and impressive segments? 4) Quality: What is the overall quality of the melody?

\section{Experimental Results}
In this section, we first compare TeleMelody with baselines to demonstrate its effectiveness. Then, we show the analysis of TeleMelody in the aspects of controllability and data efficiency.  Audio samples of the generated melodies are available via this link$\footnote{\url{https://ai-muzic.github.io/telemelody}}$ and also in Supplementary Materials as described in Appendix \ref{sec:sample}.

\begin{table*}[t!]
  \centering
  \resizebox{\linewidth}{!}{
      \begin{tabular}{l r r r |r r r r}
      \toprule
      $\empty$ & \multicolumn{3}{c}{Objective} & \multicolumn{4}{c}{Subjective} \\
      \empty & PD($\%$)$\uparrow$ & DD($\%$)$\uparrow$ & MD$\downarrow$ &Harmony$\uparrow$ & Rhythm$\uparrow$ & Structure$\uparrow$ & Quality$\uparrow$ \\ \midrule
      
      (EN) Transformer Baseline & 24.41&47.04& 2.69&2.0 ($\pm$0.18)&2.1 ($\pm$0.19)&1.8 ($\pm$0.16)&1.8 ($\pm$0.16)\\
      (EN) SongMASS & 34.32 & 47.88 & 2.61 & 2.4 ($\pm$0.20) & 2.3 ($\pm$0.20)& 2.3 ($\pm$0.20) & 2.2 ($\pm$0.19)\\
      (EN) \textbf{TeleMelody}& \textbf{43.86} & \textbf{52.06} & \textbf{2.40} &\textbf{3.2 ($\pm$0.24)} & \textbf{3.4 ($\pm$0.19)}& \textbf{3.3 ($\pm$0.21)}&\textbf{3.3 ($\pm$0.20)}\\ 
      \midrule
       
      (ZH) Transformer Baseline & 11.40& 38.03&6.75&1.5 ($\pm$0.15)&2.0 ($\pm$0.22)&1.5 ($\pm$0.14)&1.5 ($\pm$0.13)\\
    
      (ZH) SongMASS & 26.36 & 50.09& 3.29&1.8 ($\pm$0.19)& 2.2 ($\pm$0.20)& 1.8 ($\pm$0.16)& 1.7 ($\pm$0.17)\\
        (ZH) \textbf{TeleMelody}& \textbf{49.76} & \textbf{51.48} & \textbf{2.80} & \textbf{3.0 ($\pm$0.19)}& \textbf{3.5 ($\pm$0.19)}& \textbf{3.3 ($\pm$0.19)}&\textbf{3.2 ($\pm$0.19)}\\
      \bottomrule
        
      \end{tabular}
  }
  \caption{Objective and subjective evaluation (with 95\% confidence interval) of TeleMelody and the baseline systems.}
  \label{tab:result}
\end{table*}

\subsection{Main Results}
\label{sec:main results}
We compare the performance of TeleMelody with two baselines: 1) SongMASS \cite{sheng2020songmass}, the state-of-the-art system that deals with low-resource scenario by end-to-end unsupervised lyric-to-lyric and melody-to-melody pre-training; 2) Transformer baseline, a Transformer model directly trained with paired lyric-melody data. Our TeleMelody (including lyric-to-rhythm and template-to-melody models) has similar number of model parameters with that of SongMASS and Transformer baseline for fair comparison. For English, the two baselines use $8,000$ paired lyric-melody data \cite{yu2021conditional}, and SongMASS additionally uses $362,237$ unpaired lyrics and $176,581$ unpaired melodies for pre-training. For Chinese, the two baselines use $18,000$ paired lyric-melody data \cite{bao2019neural}, and SongMASS additionally uses $228,000$ unpaired lyrics and $283,000$ unpaired melodies crawled from the Web.

As shown in Table \ref{tab:result}, TeleMelody significantly outperforms Transformer baseline on all the objective metrics (Improvement on EN: $19.45\%$ in PD, $5.02\%$ in DD, and $0.29$ in MD; Improvement on ZH: $38.36\%$ in PD, $13.45\%$ in DD and $3.95$ in MD). Compared with SongMASS, TeleMelody also performs better on all the objective metrics (Improvement on EN: $9.54\%$ in PD, $4.18\%$ in DD, and $0.21$ in MD; Improvement on ZH: $23.40\%$ in PD, $1.39\%$ in DD and $0.49$ in MD). Meanwhile, for all the subjective metrics, TeleMelody is better than both the Transformer baseline and SongMASS. Specifically for the quality, TeleMelody outperforms Transformer baseline by $1.49$ in EN and $1.7$ in ZH, and outperforms SongMASS by $1.10$ in EN and $1.43$ in ZH. 

The results show that an end-to-end Transformer model has poor performance, since available paired lyric-melody data is limited. In SongMASS, the lyric-to-lyric and melody-to-melody unsupervised pre-training can effectively improve the performance of end-to-end models, but it is still insufficient since the unlabeled data is not effectively utilized to improve the correlation learning between lyrics and melodies. TeleMelody performs the best, since it successfully reduces the difficulty and effectively utilizes the unpaired melodies in the two-stage framework. Besides, the improvements in Table \ref{tab:result} are consistent with the intuition in designing the template: 1) tonality and chord progression in the template can control note pitch and thus help improve PD and harmony; 2) rhythm pattern and cadence in the template can control both onset and duration of notes, and thus help improve DD and rhythm; 3) the repetitive patterns in the template can help improve structure.

\subsection{Method Analyses}
In this subsection, we verify the effect of the designed template and the alignment regularization on controllability and verify data efficiency by testing TeleMelody with varying paired lyric-rhythm training data.


\paragraph{Controllability Study}
As shown in Table \ref{tab:alignment}, TeleMelody with template is quite high in RA and AA, and close to the ground truth in CA, which indicates that the melody generated by TeleMelody is highly consistent with the rhythm, chord, and cadence in the template. Meanwhile, TeleMelody can also control the tonality with a good TA accuracy with the template. Moreover, the proposed alignment regularization (AR) can further improve the controllability (with closer TA, CA, RA, and AA to the ground truth).

\begin{table}[th] 
 
  \centering
   \resizebox{\columnwidth}{!}{
      \begin{tabular}{l c c c c}
       \toprule
       \empty Metric
        \empty & TA($\%$)&CA($\%$) & RA($\%$) & AA($\%$)\\
       \midrule
       Ground Truth & 100.00 & 62.10  &  100.00& 100.00\\
       \midrule
       Template & 75.64  & 64.11 & 99.88 & 99.94\\
        \textbf{Template+AR} & \textbf{77.41}  & \textbf{62.62} & \textbf{99.99} &\textbf{99.98} \\
       \bottomrule
      \end{tabular}
  }
 \caption{Results on model controllability. The closer that TA, CA, RA, AA are to Ground Truth, the better the controllability is.}
  \label{tab:alignment}
\end{table}

\begin{figure}[t]
\centering
\subcaptionbox{Template.\label{fig:wo_align}}{

\includegraphics[width=0.45\columnwidth]{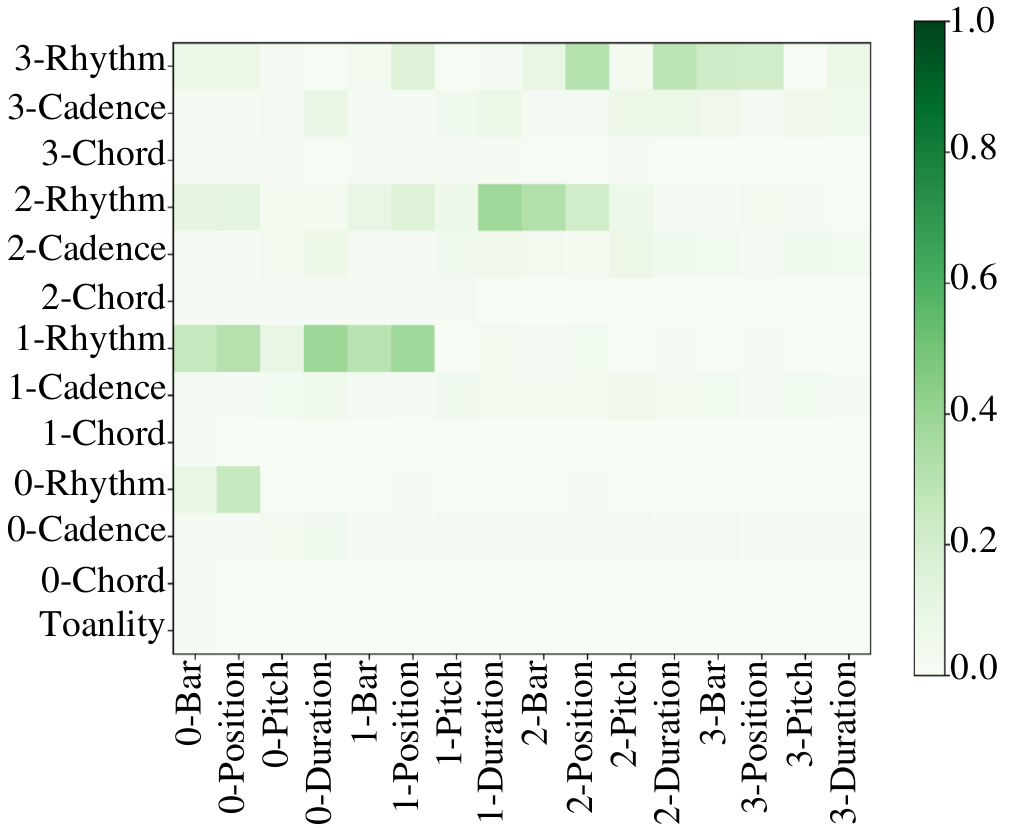}
}
\subcaptionbox{Template+AR.\label{fig:w_align}}{ 

\includegraphics[width=0.45\columnwidth]{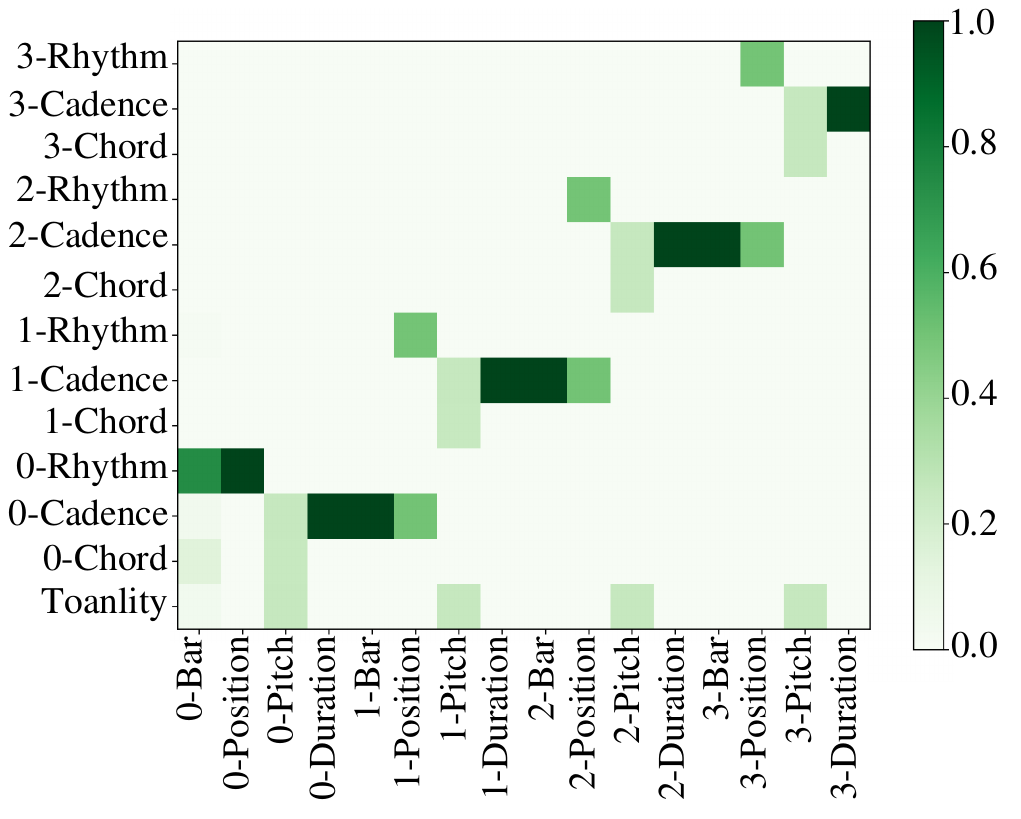}
}
\caption{Alignment visualization. We denote x-coordinate ``$i$-feature'' as the feature of the $i^{th}$ note in the melody sequence, and y-coordinate ``$i$-element'' as the musical element of $i^{th}$ note in the template sequence.}
\label{fig:comp}
\end{figure}

To show the effect of the alignment regularization intuitively, we further visualize the average encoder-decoder attention weights of all heads in the last layer. As shown in Figure \ref{fig:comp}, after adding alignment regularization (template + AR), the related elements in the template and the generated melody are clearly aligned, according to the alignment rules introduced in Section \ref{sec:align}.

\begin{figure}[t]
\centering
\subcaptionbox{Basic melody. Tonality is ``C major'' and chord progression is ``vi(Am)-IV(F)-V(G)-I(C)''.\label{fig:base}}{

\includegraphics[width=1.0\columnwidth,height=0.15\columnwidth]{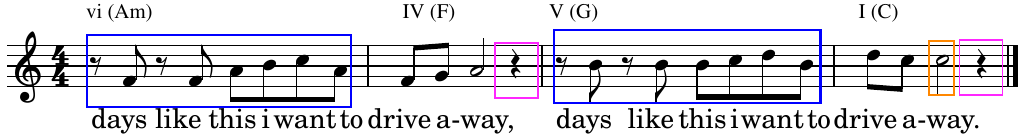}
}

\subcaptionbox{Adjusting tonality to ``A minor''.\label{fig:tonality}}{
\includegraphics[width=1.0\columnwidth,height=0.15\columnwidth]{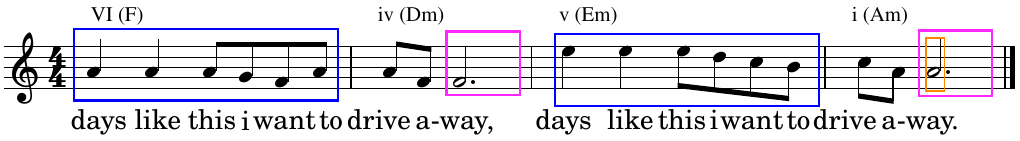}
}

\subcaptionbox{Adjusting chord progression to ``V(G)-I(C)-V(G)-I(C)''.\label{fig:chord}}{
\includegraphics[width=1.0\columnwidth,height=0.15\columnwidth]{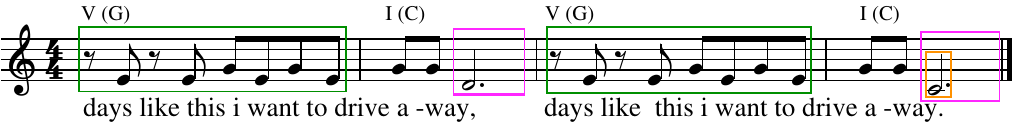}
}
\vspace{-3mm}

\caption{Case study on template adjustment. Similar bars are labeled by blue boxes while repetitive bars are labeled by green boxes.}
\label{fig:comp_control}
\end{figure}

In addition, we conduct a case study to illustrate controllability and how the elements in the template affect the generated melody. The basic melody is shown in Figure \ref{fig:base}. We evaluate the control performance from the following aspects:

\begin{itemize}

\item Tonality determines pitch distributions in the melody. To verify the control of tonality, we adjust the tonality with fixed chord progression(VI-IV-V-I) in the template. As shown in Figure \ref{fig:tonality}, when we change the tonality in the template from ``C major'' to ``A minor'', pitch distribution changes and the pitch of the ending note changes from tonic pitch of ``C major'' to tonic pitch of ``A minor''.

\item For each note, a chord is provided in the template, which affects the pitch of the note. As shown in Figure \ref{fig:chord}, when we change the chords of the notes in the first and the second bar, the pitches are changed correspondingly. 
  
\item Rhythm affects the onset position of the note. For example, in Figure \ref{fig:comp_control}, when we use the same rhythm for the first and third bars as labeled by green or blue boxes in each melody, the onset positions of the notes in the two bars are the same.

\item Cadence affects the note onset intervals and note pitch at the end of phrases in the generated melody. As shown from pink boxes in Figure \ref{fig:comp_control}, note onset intervals at the end of each sentence are larger than other places from melody when cadence of note before comma and period are labeled as “half cadence” or an “authentic cadence”. And the note pitch from the orange box is the tonic pitch when “authentic cadence” is assigned to period.

\end{itemize}

\paragraph{Data Efficiency Study}

TeleMelody is data-efficient: 1) In template-to-melody module, we train a model with extracted templates from melodies in a self-supervised way. 2) In lyric-to-template module, only the lyric-to-rhythm model requires paired training data, which is easier to obtain than human-labeled paired lyric-melody data.

Since lyric-rhythm data is the only paired data we use, we test the lyric-to-rhythm model with different kinds of lyric-rhythm data to demonstrate the data-efficiency performance on TeleMelody: 
\begin{enumerate}
    \item \textbf{human-labeled data:} lyric-rhythm data obtained from the human-labeled paired lyric-melody data utilized by baselines described in Section \ref{sec:main results}.
    \item \textbf{100\% crawled data:} lyric-rhythm data obtained from crawled  lyric-audio data described in Section \ref{sec:dataset}, which is of comparable size with human-labeled data in 1.
    \item \textbf{50\% crawled data:} the same as 2 except only 50\% of the crawled lyric-audio data is used.
    \item \textbf{w.o. data:} no paired lyric-rhythm data is used, since we replace the lyric-to-rhythm model with hand-craft rules. The details of these hand-craft rules are described in Appendix \ref{sec:l2r rule}.
\end{enumerate}

It is shown in Table \ref{tab:scale} that: 
\begin{itemize}
    \item Comparing TeleMelody that uses crawled data with TeleMelody that uses human-labeled data, there is no distinct drop in performance, which shows the data efficiency of TeleMelody in that it can achieve good performance without any human-labeled data.
    \item Comparing TeleMelody that uses 100\% of the crawled data with TeleMelody that uses only 50\% of the crawled data, the performance only declines slightly, while still outperforms SongMASS on all the metrics. 
    \item To further demonstrate the efficiency on the lyric-to-rhythm model, we replace the lyric-to-rhythm model with hand-craft rules for extracting rhythm pattern from lyrics, so that no paired data is used. The performance of TeleMelody without the proposed lyric-rhythm model significantly degrades, which demonstrates the advantage of the lyric-to-rhythm model in TeleMelody.
    \item Comparing with SongMASS that uses human-labeled lyric-melody data, all the above settings of TeleMelody achieves better performance on all the objective metrics in English. These promising results further illustrate the data efficiency of our proposed TeleMelody. 
\end{itemize}

\begin{table}[t]\small
  \centering
  \resizebox{\linewidth}{!}{
  \begin{tabular}{l r r r}
  \toprule
  \empty & PD($\%$)$\uparrow$ & DD($\%$)$\uparrow$ & MD$\downarrow$ \\ \midrule 
  
    (EN) TeleMelody &&&\\
    \qquad\quad$\bullet$ human-labeled data& \textbf{46.73} & \textbf{52.95}  & \textbf{2.36} \\
    \qquad\quad$\bullet$ $100\%$ crawled data& 43.86& 52.06 & 2.40 \\
  \qquad\quad$\bullet$ $50\%$ crawled data &41.42 & 48.59 & 2.40 \\
 \qquad\quad$\bullet$ w.o. data & 35.60 & 47.93 & 2.55\\ 
  (EN) SongMASS & 34.32 & 47.88 & 2.61 \\
    \midrule
 (ZH) TeleMelody &&&\\
 \qquad\quad$\bullet$ human-labeled data& 47.29 & \textbf{54.76} & \textbf{2.24}\\
  \qquad\quad$\bullet$ $100\%$ crawled data & \textbf{49.76}& 51.48 & 2.80 \\
  \qquad\quad$\bullet$ $50\%$ crawled data&46.25 & 50.14 & 3.12 \\ 
  \qquad\quad$\bullet$ w.o. data & 30.22 & 33.84 & 4.95\\ 
  (ZH) SongMASS & 26.36 & 50.09& 3.29\\
 
  \bottomrule
    
  \end{tabular}
  }
  \caption{Objective evaluations of systems with varying lyric-rhythm training data. The ``human-labeled data'' is obtained from the human-labeled paired lyric-melody data utilized by baselines described in Section \ref{sec:main results}, while the ``crawled data'' is obtained from crawled singing-lyric data described in Section \ref{sec:dataset}. We denote ``w.o. data'' as replacing the model with hand-craft rules described in Appendix \ref{sec:l2r rule}.}
  \label{tab:scale}
\end{table}

\section{Conclusion}
In this paper, we proposed TeleMelody, a two-stage lyric-to-melody generation system with music template (e.g., tonality, chord progression, rhythm pattern, and cadence) to bridge the gap between lyrics and melodies. TeleMelody is data efficient and can be controlled by end users by adjusting the musical elements. Both subjective and objective experimental evaluations demonstrate that TeleMelody can generate melodies with higher quality than previous lyric-to-melody generation systems. Experimental studies also verify the data efficiency and controllability of TeleMelody. In future work, we will extend our proposed two-stage framework to other music generation tasks (e.g., melody-to-lyric generation and melody-to-accompaniment generation).

\section{Limitations}
With the use of the template to connect lyrics with melodies, TeleMelody can outperform the state of the art, together with less requirement on paired training data. Nevertheless, we still find that the generated melodies are relatively less diverse than human creations. Future work can further improve this in the following aspects: 1) The large-scale model is demonstrated to be effective for learning the underlying data distribution, which can help improve the diversity of generated melodies, but it requires large amount of paired data.  Thus, more works on large-scale data annotation and data collection techniques (e.g. accurate audio-to-MIDI conversion) are expected to be developed. 2) Experimental results demonstrate the introduction of the template containing well-designed musical elements in TeleMelody helps address the issue of low-resource data. More musical elements in the template are expected to be extended to other scenarios (e.g. controlling of music style and emotion).

\bibliography{anthology,custom}
\bibliographystyle{acl_natbib}

\appendix

\section{Pipeline of Collecting Lyric-Rhythm Data }
\label{sec:l2r}
We crawl paired lyric-singing audio data from the Web and then utilize spleeter$\footnote{\url{https://github.com/deezer/spleeter}}$, a public music separation tool, to separate the vocal from the accompaniment part. The timestamps of lyrics and the tempo of melodies are two necessary information during collection of rhythm information. To extract lyric timestamps, we first split the singing audio into sentence-level segments with crawled start and end timestamp. We convert lyrics into sequences of phonemes via Phonemizer$\footnote{\url{https://github.com/bootphon/phonemizer}}$ and then obtain the vocal-lyric alignment in phoneme level with an audio alignment tool, Montreal Forced Aligner$\footnote{\url{https://github.com/MontrealCorpusTools/Montreal-Forced-Aligner}}$. Based on these phoneme-level vocal-lyric alignments, we can obtain the corresponding timestamp of each lyric. To extract tempo information, we perform a direct estimation from the accompaniment audio with an audio information retrieval tool, librosa$\footnote{\url{https://github.com/librosa/librosa}}$. Finally, with tempo information and timestamp of each lyric, we can infer beat-level onset, that is, the corresponding rhythm of each lyric.  

Considering that syllables in lyrics correspond to
notes in melodies, we encode each syllable as a lyric token.
For Chinese, each character has only one syllable. And for
English, we divide each English word into a number of syllables, and represent each syllable as a lyric token.

\section{Processing Details of Melody and Template}
\label{sec:process}

\paragraph{Melody}
In this paper, we only consider the melody with a constant tempo and a time signature of $4/4\footnote{4/4 denotes that that each beat is a 1/4 note and each bar has 4 beats.}$. Each note is represented by four consecutive tokens (bar, position, pitch and duration). We use $256$ tokens to represent different bars and $16$ tokens to represent different positions in a bar with granularity of $1/16$ note. We use $128$ tokens to represent pitch values following the MIDI format. We use $16$ tokens to represent duration values ranging from a $1/16$ note to a whole note. 

We perform pre-processing to get high-quality melody data as follows: First, we extract the melody from the track$\footnote{\url{https://en.wikipedia.org/wiki/Multitrack\_recording}}$ with at least $50$ notes that has the highest average pitch among all the tracks, and then delete polyphonic notes. Second, we normalize the tonality to ``C major'' or ``A minor'', and normalize the pitches to fit the range of vocals. Finally, we filter the empty bars in each sample. After these steps, we can obtain the processed target melody dataset. To construct paired template-melody dataset, we utilize our proposed knowledge-based rules to extract the corresponding template from the target melody.

\paragraph{Template}
In this paper, template contains musical elements including tonality, chord progression, rhythm pattern, and cadence. We only consider ``C major'' and ``A minor'' tonalities for simplicity, since other tonalities can be transposed to these two tonalities based on their scales. Chord consists of a root note and a chord quality. We consider $12$ chord roots ($C,C\sharp{},D,D\sharp{},E,F,F\sharp{},G,G\sharp{},A,A\sharp{},B$) and $7$ chord qualities (major, minor, diminished, augmented, major$7$, minor$7$, half diminished), resulting in $84$ possible chords in total. We use $4$ tokens ranging from $0$ to $3$ to represent rhythm patterns, that is, beat-level onset position in a bar. For cadence, we consider ``half cadence'', ``authentic cadence'' and ``no cadence'', which is aligned with comma, period and other syllables in lyrics respectively.

\section{Dataset Statistics}
\label{sec:data_stat}
Dataset statistics are shown in Table \ref{tab:l2r} and \ref{tab:t2m}.

\begin{table}[ht]
  \centering
  
  \begin{subtable}{\columnwidth}\centering
  {
  \resizebox{\columnwidth}{!}{
  \begin{tabular}{l r}
        \toprule
       \# of data samples & $9,761$ \\
       Average \# of words per song & $26.02$ \\
       Average \# of syllables per song & $31.15$ \\
       Average \# of punctuation marks per song & $3.79$  \\
       \bottomrule
      \end{tabular}
     } 
     }
    \caption{English}
  \end{subtable}
  \hfill
  \begin{subtable}{\columnwidth}\centering
  {
  \resizebox{\columnwidth}{!}{
  \begin{tabular}{l r}
   \toprule
   \# of data samples & $74,328$ \\
   Average \# of words per song & $78.77$ \\
   Average \# of punctuation marks per song & $9.30$  \\
   \bottomrule
    \end{tabular}
    }
    }
   \caption{Chinese}
  \end{subtable}
 \caption{Statistics of lyric-rhythm dataset.}
 \label{tab:l2r}
  
\end{table}

\begin{table}[ht]
  \centering
  
  \begin{tabular}{l r}
   \toprule
   \# of data samples  & 157,702 \\
   Average \# of notes per song & 28.76 \\
   Average \# of bars per song & 5.93\\
   \bottomrule
  \end{tabular}
  \caption{Statistics of template-melody dataset.}
  \label{tab:t2m}

\end{table}



\section{Cadence Extraction Rule}
\label{sec:cad}

In template-to-melody module, we extract cadence from melodies in training stage through cadence extraction rule as follows : 1) ``no cadence'' is assigned to the note when the note duration is short (e.g., less than $1$ beat$\footnote{\url{https://en.wikipedia.org/wiki/Beat\_(music)}}$) or the onset interval between the current note with the next note is small (e.g., less than $1.5$ beats).  2) ``authentic cadence'' is assigned to the note when the note is the root note$\footnote{\url{https://en.wikipedia.org/wiki/Root\_(chord)}}$ of tonic$\footnote{\url{https://en.wikipedia.org/wiki/Tonic\_(music)}}$ chord, or the inferred chord of this note is tonic chord. There is also a probability of $p$ (e.g., 0.3) for labeling this note with ``authentic cadence'' when the note is other notes in tonic chord rather than root note, and the onset interval is large (e.g., more than $2$ beats). 3) ``half cadence'' is assigned to notes outside the above two situations. In lyric-to-template module, we directly obtain cadence from lyrics in inference stage through punctuation-to-cadence mapping.

Therefore, a question may arise: is there a gap in cadence between training and inference? To answer this question, we explore the statistics of cadences in template-melody dataset. The results are shown in Table \ref{tab:cadence}:
\begin{itemize}
  \item Notes labeled with ``no cadence'', which are aligned with syllables in lyrics, have short duration and onset interval. Notes labeled with  ``half cadence'', which are aligned with commas in lyrics, have $4.91\times$ longer duration and $4.22\times$ longer onset interval than those labeled with ``no cadence''. Notes labeled with  ''authentic cadence'', which are aligned with periods in lyrics, have $5.94\times$ loner duration and $5.11\times$ longer onset interval than those labeled with ``no cadence''. This is in consistent with musical knowledge, since punctuation marks are usually aligned with pauses.
  
  \item The average number of ``no cadence'' is $5.86\times$ greater  than the average number of ``half cadence' and ``authentic cadence'' combined. This ratio is similar to the ratio of syllables to punctuation marks in lyrics, as shown in Table \ref{tab:l2r}. 
\end{itemize}

\begin{table}[ht]\small
  \centering
  \resizebox{\columnwidth}{!}{
  \begin{tabular}{l r r r}
   \toprule
  Cadence & Average \# per song & Duration & Onset Interval \\
   \midrule
   No  & 24.57 & 1.74 & 2.29 \\
   Half & 2.53 & 8.54 & 9.66 \\
   Authentic & 1.66 & 10.34 & 11.70 \\
   \bottomrule
  \end{tabular}
  }
 \caption{Statistics of notes labeled with different cadences in template-to-melody dataset. Both duration and onset interval are quantified to $1/16$ note.}
   \label{tab:cadence}
 
\end{table}

\section{Lyric-to-Rhythm Hand-Craft Rules}
\label{sec:l2r rule}
We design several lyric-to-rhythm rules to demonstrate that TeleMelody can be applied in the scenario without any paired data. 
Specifically, for English lyrics, a note is corresponding to a syllable, and we generate the rhythm patterns syllable by syllable, where the onset interval between a note and its previous note is $2$ beats if its corresponding syllable is the start of a sentence, is $1$ beat if its corresponding syllable is the start of a word but not the start of a sentence, and is $0.5$ beat otherwise. For Chinese lyrics, a note is corresponding to a character, and we generate the rhythm patterns character by character, where the onset interval between a note and its previous note is $2$ beats if its corresponding character is the start of a sentence, and is $1$ beat otherwise. 

\section{Definitions of TA, RA, CA, and AA}
\label{sec:def}
We evaluate model controllability with accuracy of tonality, rhythm pattern, chord, and cadence (TA, RA, CA, AA), that is, the proportion of notes in consistent with given template. Specifically, a melody is in consistent with tonality if the inferred tonality is the same as the template tonality; a note is in consistent with rhythm pattern if its position is in accord with given rhythm pattern information; a note is in consistent with chord if its pitch is within the chord; a note is in consistent with cadence if both its duration and its onset interval between this note and the next note comply with the extraction rules described in Section \ref{sec:retrieval}. 


\section{Samples of Generated Melodies}
\label{sec:sample}
Samples of generated melodies are available via this link$\footnote{\url{https://ai-muzic.github.io/telemelody}}$  and are available in ``sample.zip''. We utilize XStudioSinger$\footnote{\url{https://singer.xiaoice.com}}$ tool to synthesise singing  and ChordPulse$\footnote{\url{http://www.chordpulse.com}}$ for accompaniment for better listening experience.

\end{document}